\documentclass[12pt]{iopart}
\usepackage{graphicx}
\begin{document}
\title{Theory for stability and regulation of epigenetic landscapes}

\author{Mille A. Micheelsen,* 
Namiko Mitarai,* \\
Kim Sneppen,*\thanks{Corresponding author.  E-mail address: sneppen@nbi.dk}
and
Ian. B. Dodd.$\dagger$}
\address{\
*Center for Models of Life, Niels Bohr Institute,
Blegdamsvej 17, DK-2100, Copenhagen O, Denmark.\\
$\dagger$Department of
Molecular and Biomedical Sciences (Biochemistry), University of
Adelaide SA 5005, Australia.
}

\begin{abstract}
Cells can often choose among several stably heritable phenotypes.
Examples are the expression of genes in
eukaryotic cells where long chromosomal regions can adopt persistent
and heritable silenced or active states,
that may be
associated with positive feedback in dynamic modification of nucleosomes. 
We generalize this mechanism in terms
of bistability associated with valleys in an epigenetic landscape.
A transfer matrix method was used to rigorously follow the system through
the disruptive process of
 cell division.
This combined treatment of noisy dynamics both between and during cell
division provides an efficient way to calculate the stability of alternative
states in a broad range of epigenetic systems.

\emph{Key words:} epigenetic landscape; nucleosome modification; bistability;
positive feedback 
\end{abstract}
\pacs{87.18.-h,87.10.Mn,87.17.-d}

\section*{Introduction}
Cells carry information handed down from their
ancestors and are able to pass on information to their descendants.
In many cases this ``memory'' is epigenetic, that is,
not stored in the DNA sequence, allowing cells with identical
DNA to maintain distinct functional identities.
Epigenetic cell memory implies alternative states of gene expression
that are stable over time and are inherited through cell division.

A proposed mechanism for epigenetic cell memory invokes positive
feedback loops in nucleosome modification
\cite{grunstein,turner,felsenfeld,dodd, sedighi,sneppen08}.
Positive feedback is a mechanism seen in many
other regulatory systems where for example production of
a regulatory protein activates its own production, or more robustly
where two mutual repressors act strongly enough to prevent co-expression. 
A complementary view on cell memory is that of an epigenetic landscape 
\cite{waddington,goldberg}, where the
state of a cell develops on some potential energy surface, and a
state is maintained when the cell is caught at a particular valley
for a long time.

In this paper we develop an epigenetic landscape formalism for cell
memory by positive feedback in nucleosome modification.
Instead of viewing cell differentiation as the 'pushing' of a cell
over a fixed landscape \cite{waddington,goldberg}, 
our approach suggests that cell fate could be controlled 
by changing the landscape.
\section*{Models}
Inspired by the mating type switch in {\it S. pombe} \cite{thon} we
introduced a model for bistability by positive feedback in
nucleosome modification \cite{dodd}. The model had one parameter,
the positive feedback to noise ratio 
$F$, 
and modeled the dynamics of a system
consisting of $N$ nucleosomes where each could be in one of
3 states, modified, unmodified and anti-modified.

\begin{figure}[!h]
\includegraphics[width=0.65\textwidth]{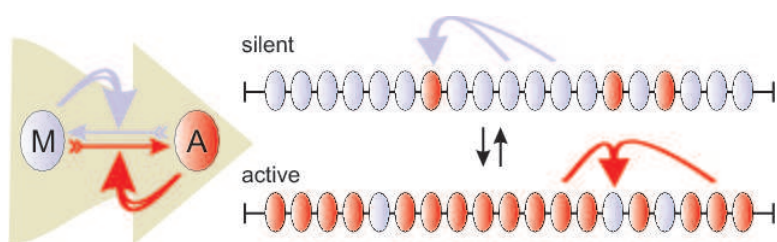}
\caption{{\bf Model.} Consider a 
system with $N$ sites. Each site represents a nucleosome that can
be modified (M) or anti-modified (A). Transitions between these two
states are in part random, and in part auto-regulated by recruitment
of histone modifying enzymes by local nucleosomes: At each update, a
nucleosome $i$ is either, with probability
$1-\alpha$, set to an $M$ or an
$A$ state randomly. Or with probability
$\alpha$, two other nucleosomes are chosen, and if
these are in the same state then the state of nucleosome
$i$ is set to this state. 
The model is parametrized by 
the positive feedback to noise ratio $F=\alpha/(1-\alpha)$.
\label{fig1} }
\end{figure}

Here we introduce a simpler version of this modification system, in which there are only two
chemical states of each nucleosome. We term these modified (M) and anti-modified (A) to
indicate their mutual exclusivity, with the A nucleosome carrying either a different chemical
modification or no modification (Fig.~\ref{fig1}). 
Each nucleosome type recruits a modifying enzyme
that converts the other type to its own type. 
Our results from \cite{dodd} demonstrated that robust bistability requires
an effective cooperativity in the recruitment process.
Cooperativity could here be included directly by requiring
that 2 local nucleosomes with the same modification, e.g.  
M, are needed to make an 
A$\rightarrow $M conversion, as described in Fig.~\ref{fig1}.
Note that this model is not only just a simplified version 
of 3-state model but also 
has parallels with 
the mating-type silencing system in {\it S. cerevisiae}, 
where one typically considers acetylated
and non-acetylated nucleosomes \cite{rusche,naar,ruthenburg}.

Suppose $M$ nucleosomes are in the M-state at time $t$.
We now express the development of the fraction $m=M/N$ 
sites in the M-state.
Denoting $dm=1/N$, we have
\begin{equation}
\frac{dm}{dt}  =  \left(R_+(m) - R_-(m)\right)\cdot dm  + noise  \label{eqV0}
\end{equation}
with the rate that
the system with fraction $m$ of M-sites 
gets one more (or one less) M-sites being $R_{+}(m)$ (or $R_-(m)$) and with
$noise$ having zero mean and being 
associated with the randomness of processes in a finite system. 
The rates are given by
\begin{eqnarray}
R_+(m) &=& \alpha (1-m)m^2 + (1-\alpha)(1-m), \nonumber \\
R_-(m) &=& \alpha m(1-m)^2 + (1-\alpha)m.
\label{eqPflux}
\end{eqnarray}
Here, 
the first term is the nucleosome recruitment;
in the case of $R_+(m)$,
the recruitment occurs with probability $\alpha$, and 
it must involve 2 M-sites (probability proportional to $m^2$)
and must change the modification on an A-site (probability  proportional to
$(1-m)$). 
The second term 
is the noise effect which is proportional to $(1-\alpha)$,
where a nucleosome can become M by random conversion of from an A state 
(probability proportional to $(1-m)$), and {\it vice versa}.
This noise represents all events from the cell that are not associated to
the direct recruitment processes from other nucleosomes within our N
nucleosome system. One should be aware that there is another level of
noise in our stochastic description, represented by the {\sl noise} 
in eq.(\ref{eqV0}) which is the noise associated with 
the stochasticity of the molecular processes.

The ratio of recruitment (or the positive feedback) to the 
noise, $F=\alpha/(1-\alpha)$ is the parameter of the model.
The Langevin equation for $m$ is then
\begin{eqnarray}
\frac{dm}{dt} & = &  \frac{\alpha}{N} \; (2m-1)[m(1-m) - 1/F]  + noise, 
\label{eqLV}
\end{eqnarray}
implicitly implying that there is bistability when $F>4$ provided that the $noise$ term 
is small (system size large).

To analyze the time development of the distribution of $m$ more carefully we 
reformulate the model in terms of a master equation for the probability 
$P(m,t)$ as
\begin{eqnarray}
\label{eqdP}
\frac{\partial }{\partial t} P(m,t)
& = & 
R_-(m+dm)P(m+dm,t) + 
R_+(m-dm)P(m-dm,t)\nonumber \\
&&
- [R_+(m)+R_-(m)]P(m,t).
\end{eqnarray}

\section*{Results}
\subsection*{Epigenetic landscape generated from a positive
feedback system} 
We extract ``the potential landscape'' in $m$-space
by comparing eq.~(\ref{eqdP}) with the
generic 1-dimensional Fokker-Planck equation for diffusion of 
a particle in a potential
$U(m)$: 
\begin{eqnarray}
\frac{\partial }{\partial t}P  & = &  - \frac{\partial J}{\partial m} =
-\frac{\partial }{\partial m} \left[-\mu(m) \frac{d U(m)}{dm} P
- \frac{\partial (D(m)P)}{\partial m}\right] \nonumber
\\
& = & -\frac{\partial }{\partial m} \left[-\mu(m) \frac{dV(m)}{dm} P
- D(m)\frac{\partial P}{\partial m}\right] 
\label{eqFokkerPlanck}
 \end{eqnarray}
Here, $\mu (m)$ is the mobility
and $D(m)$ quantifies the stochastic motion in terms
of an $m$-dependent diffusion coefficient.
In the last step, $V(m)$ represents an
effective potential that includes both drift and noise events,
defined as $\frac{dV}{dm}=\frac{dU}{dm} + \frac{D}{\mu }
\frac{d \ln(D)}{dm}$. 

Expanding eq.~(\ref{eqPflux}) with eq.~(\ref{eqdP}) to second order
in $dm=1/N$ and comparing it with 
eq.~(\ref{eqFokkerPlanck}),
we find the drift 
$\left\langle \frac{dm}{dt} \right\rangle=\mu(m)(dU/dm)$
, the effective potential $V(m)$, 
diffusion $D(m)$ and mobility $\mu (m)$ as the following:
\begin{eqnarray}
\left\langle \frac{dm}{dt} \right\rangle  & = & \frac{\alpha}{N} \; (2m-1)[m(1-m) - 1/F] \label{eqV1}\\
     V(m) & = & 2N m(1-m) + \left(1-\frac{4N}{F}\right)
\ln[Fm(1-m)+1] \label{eqV}\\
D(m) & = & \mu(m) = \alpha \; \frac{m (1-m) + 1/F }{2 N^2}. \label{diff}
\end{eqnarray}
Here, the first equation could have been obtained directly from the
Langevin equation (\ref{eqLV}).
From these expressions we again can see 
that there is the critical recruitment to noise
ratio $F=\alpha/(1-\alpha)$, with $m=1/2$ being an unstable fixed point for $F>4$.

\begin{figure}[!ht]
\includegraphics[width=0.65\textwidth]{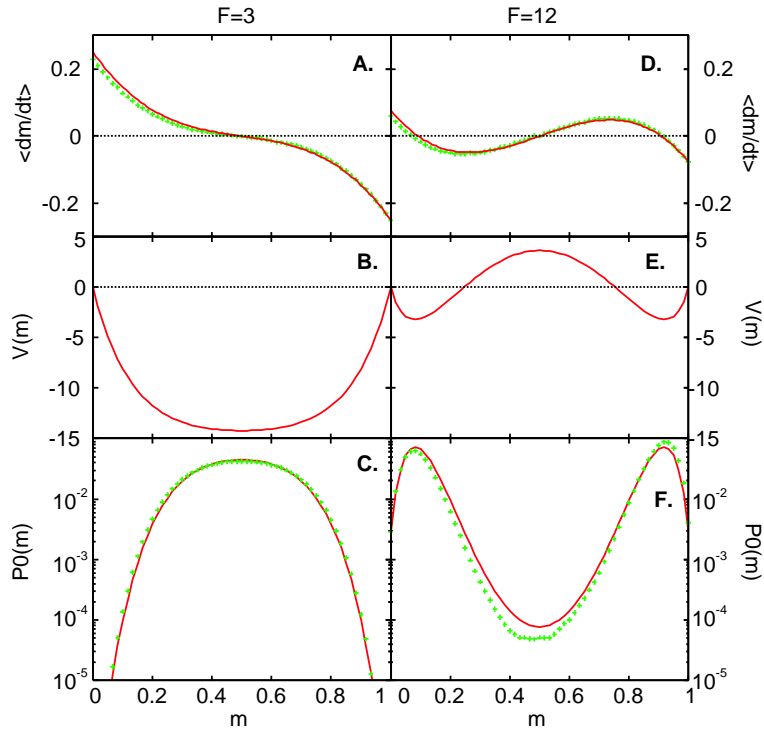}
\caption{{\bf System properties.}
Analytical results (solid lines) for a system of
size $N=60$ showing; (A,D): the drift $\langle dm/dt \rangle$,
(B,E): the effective potential $V(m)$ and (C,F): the steady state
distribution $P_0(m)$ in
the two regimes: (A,B,C) - $F=3$ where there is no
bistability and (D,E,F) - $F=12$ where there is well
defined bistability, which can be seen in the 
effective potential (E) with two valleys separated by a hill. 
For $\langle dm/dt \rangle$ and $P_0(m)$ (A,C,D,F),
the numerical results are also shown by symbols.
Comparing the $V(m)$ (B,E) and $P_0(m)$ (C,F) notice that
$\ln(P_0(m)) \sim -V(m)$ as expected in the steady state from
requiring that $J=0$ in eq.~(\ref{ss}).
\label{fig3} }
\end{figure}

\begin{figure}[!h]
\includegraphics[width=0.65\textwidth]{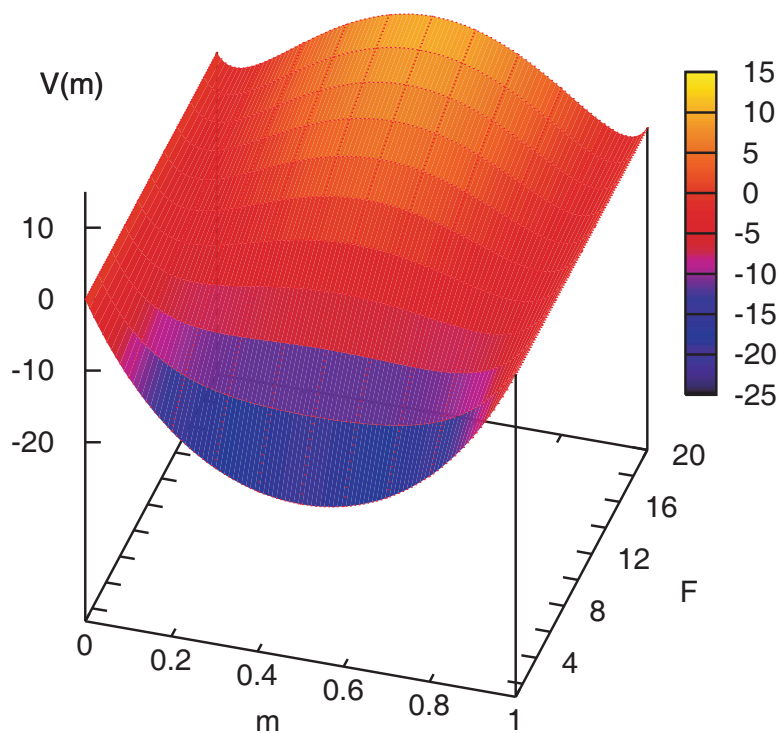}
\caption{{\bf An epigenetic landscape generated from a positive
feedback system.} 
Here the effective potential $V(m)$ from eq.~(\ref{eqV}) 
is plotted as a function of $F$
with fixed $N=60$. 
The landscape changes gradually
as $F$ increases, from a single steep valley,
through an almost equipotential 'river plain', to two valleys.
This change is associated with stronger recruitment
processes at larger $F$ values.
\label{fig2} }
\end{figure}

Figure~\ref{fig3} 
shows $\langle dm/dt\rangle$, $V(m)$ and
the steady state distribution $P_0(m)$ for $F=3$ and
$F=12$, thereby illustrating mono-stable and bistable systems. 
Also note that the analytic results fit the stochastic simulation, 
with a deviation that scales as $1/N$ with increased
system size (not shown).
Figure~\ref{fig2} shows how the epigenetic landscape changes gradually
as  $F$ increases: from a single steep valley,
through an almost equipotential 'river plain', to two valleys.
Because $F$ depends on protein concentrations and affinities, the shape of the epigenetic landscape is under biological control.

One can repeat these calculations for a model where recruitment is
not requiring the cooperative action of two nucleosomes
(The second order terms in eq.~(\ref{eqPflux})
should then be replaced by first order).
In that case one never obtains more than one stable fixed point,
confirming that bistability indeed requires cooperativity \cite{dodd,david-rus}.

The potential $V(m)$ effectively describes the effective force on $m$ 
from the combined effect of recruitment and noise events. Thus a large
positive gradient in $V(m)$ means that nucleosomes in average will tend to
lose their $m$ modification. A potential minimum, on the other hand, means
that recruitment processes and noise events balance such that the number of 
modified states typically stays around this minimum. In this way our
potential $V(m)$ plays the role of a epigenetic landscape in the Waddington
sense \cite{waddington}. In particular, the valleys and hills of this landscape can be
viewed
as the metastable epigenetic states and the barriers 
between them. We will use this analogy to calculate first the probability 
for stochastic switching between such states, and subsequently we will
discuss how one may alter the landscape by modifying the recruitment
processes that define the landscapes. 
A modification that was also envisioned by
strings in the Waddington landscape\cite{waddington}.

\subsection*{Stability of a macroscopic state}
Now we quantify the stability of a macroscopic state by the average
number of attempted updates per nucleosome before the full system
switches for the first time to the alternate epigenetic state. Using
{\footnotesize $D=\mu$} from eq.~(\ref{diff}), we in analogy to Kramers
\cite{kramer} rewrite 

\begin{eqnarray}\label{ss}
J = - D \left[ \frac{dV}{dm} P +
 \frac{\partial P}{\partial m} \right] =
 - D  \exp(-V) \frac{\partial }{\partial m} \left[P \exp(V) \right]
\end{eqnarray}
and use the quasi-stationary approximation 
(i.e., the current $J$ is constant) 
to write the
flux for going from an A-state (the potential minimum at
$m=m_A\sim 0$) to an M-state (the potential minimum at $m=m_M\sim
1$):
\begin{eqnarray}
J &=& \frac{\left[P\exp(V) \right]_{m_A}^{m_M}}{
\int_{m_A}^{m_M} (1/D(m)) \exp(V(m)) dm}.
\label{Jquasi}
\end{eqnarray}
Using a Gaussian approximation 
(i.e. $V(m)$ harmonic around both the initial
state A and the transition state T with $m=m_T = 1/2$
and the initial distribution for $P(m,t)$ around the state A), we obtain the
average life time of an epigenetic state $\tau$ as
\begin{equation}\label{Kramer}
\tau\equiv \frac{1}{|J|}  \approx  4\pi N\sqrt{\frac{4}{F}}\exp[V_T-V_A]
\end{equation}
for large $N$ and $F$, where 
$V_T=V(m=1/2)$, and $V_A=V(m_A)$ is the potential minimum 
for the A state (The detailed calculation is given in the Appendix). 
Figure~\ref{fig4} demonstrates that 
eq.~(\ref{Kramer}) 
reproduces stochastic simulations. 
However, when pushing towards very small N, there is a tendency that the 
continuous description deviate from the stochastic result. Thus,
for $N$ of order 10 or below, we recommend a stochastic simulation.

Equation~(\ref{Kramer}) can also be used to obtain an interesting prediction
from our model. Using the expression for the potential $V$ from eq.~(\ref{eqV}) 
for large $N$, we see that $V(m=\frac{1}{2}) \sim N f(F)$ 
with a function $f(F)$ independent of $N$, and thus that  
stability scales exponentially with $N$, i.e., $\tau\propto N e^{N f(F)}$.

\subsection*{Effect of cell divisions}
Epigenetic states 
are capable of being
inherited across cell divisions. 
This can give difficulties for stability of the states
\cite{dodd},
particularly for 2-state systems \cite{david-rus}
At cell
division the genome is duplicated, and following refs. \cite{dodd,annunziato}
we assume that the resident nucleosomes are partitioned randomly between 
the daughter strands. The
vacant positions are filled by new randomly selected nucleosomes where half are in M and half in A-state.
We accordingly supplement our model above with cell divisions at certain fixed time intervals. This cell generation time is measured in units
of the number of attempted nucleosome updates per nucleosome.

Whereas the potential landscape between cell divisions
drives the system toward one of the epigenetic states,
the randomization at cell divisions 
brings 
the system
closer to the top of the potential barrier in the epigenetic landscape.

Consider that before cell division the system is in a state with 
$M_b=m_b\times N$ nucleosomes in the M-state 
and the remaining nucleosomes in the A state. 
Cell division results in
the distribution of number of M-state nucleosomes $M_a$

\begin{equation}
D(M_a,M_b) = \sum_M \sum_A  \; \frac{\left(
\begin{matrix}
M_b \\
M
\end{matrix} \right)
\left( \begin{matrix} N-M_b \\
A \end{matrix} \right) \left(\begin{matrix}N-M-A\\ M_a-M\end{matrix}
\right)}{ 2^{(2N-M-A)} } \label{celldiv}
\end{equation}

\noindent where the sum runs over all the ways of getting from $M_b$
to $M_a$ by selecting $M\leq \min(M_a,M_b)$
nucleosomes in the M-state and $A\leq \min(N-M_a,N-M_b)$
nucleosomes in the A state to be transferred directly at the cell
division.

Between cell divisions, the system evolves by a stochastic sequence
of single nucleosome exchanges that can be described by motion in
the epigenetic landscape. The stochastic change 
$M\rightarrow M\pm 1$ can be followed by the master equation
above, and the system evolves towards a steady state distribution
with two well separated peaks, see Fig.~\ref{fig3}.

To combine the gradual development in a well defined epigenetic
landscape between cell division, with the sudden reshuffling at cell
divisions we express the gradual development in terms of matrix
operations. The matrix $G$ that correspond to eq.~(\ref{eqPflux}) 
is only non-zero at the diagonal and off-diagonal
elements: $G(M,M+1)=R_+(m)$, $G(M,M-1)=R_-(m)$ and
$G(M,M)=1-G(M,M-1)-G(M,M+1)$ where
$m=M/N$ and $R_\pm(m)$ is from eq.~(\ref{eqPflux}). 
The probability distribution evolves according to
$P(M)\rightarrow P(M')=\sum_{M} G(M,M')P(M)$ for
each update of a nucleosome in the system.

In Fig.~\ref{matrix} we show the time evolution of the probability
distribution from one cell division to the next for
a system with $N=60$, $F=10$ and $m\sim 0$. 
In panel A, the simulation uses a generation length of 20 updates
per nucleosome. The evolution starts just after a cell division,
where a randomization (using eq.~\ref{celldiv}) 
is followed by a drift imposed
by the epigenetic landscape. Just before the next cell division, one
resets $P(M)=0$ for $M>N/2$, and renormalizes the distribution. 
We see that at cell division a small fraction of cells reach large $m$ values, 
and over the next
$\sim 10$ updates per nucleosome can move to
$m\sim 1$. After around 10 updates, the
$P(m)$ distribution reaches a quasi-steady state, reflecting
that from then on a very small flux goes
over the barrier. Thus after 10 updates the
likelihood of further transitions between the two epigenetic states
can be ignored (for $F>10$).

The time evolution in Fig.~\ref{matrix} also illustrates
that transitions are entirely dominated by the noise at cell divisions,
at least for large enough $F$.
In fact by stochastic simulation we have verified that
switches only occur when the stochastic partitioning in a division
brings the system close to the transition state, $m\sim 1/2$.

An entire cell generation with cell division is described by the
matrix
\begin{equation}
C(M,M'')=\sum_{M'}\; G^{ (g \cdot N) }(M,M')
D(M',M'')
\end{equation}
where $g$ is the number of single
nucleosome updates per cell generation. 
Iterating the updating process 
with renormalization of the distribution, 
we obtain the probability distribution $P(M')$ shortly before
a cell division, and estimate the average number of generations 
needed before escape (escape time) 
as $n_e=\left(\sum_{M'>N/2} P(M')\right)^{-1}$
in the panel B. This escape time is the average time it takes for a system to
switch from one state to another.
From Figure \ref{matrix} we see that when $g>10$, the escape time 
does not depend on the value of $g$ for $F>10$
because of the small escaping rate after 10 updates.
%

\subsection*{Epigenetic landscape with regulated tilt}
Finally, we consider the case where the modification and anti-modification
are not symmetric but one of the effects is stronger than the other.
This asymmetry of modification could be under biological control in a real
system by changing the concentration of modifying enzymes
or by recruiting such enzymes by transcription factors \cite{sneppen08}.

One of the simplest ways to include such 
asymmetry is
to modify the transition rate eqs.~(\ref{eqPflux}) into 
\begin{eqnarray}
\label{eqPfluxtilt}
R_+(m) &=& 2 (1-\eta)\alpha (1-m)m^2 + (1-\alpha)(1-m), \nonumber \\
R_-(m) &=& 2 \eta\alpha m(1-m)^2 + (1-\alpha)m.
\end{eqnarray}
Here, a new parameter $\eta$, defined in the range $0<\eta<1$, 
sets the relative
strength of modification versus anti-modification,
and gives the symmetric case when $\eta=1/2$.

Figure~\ref{matrixx} shows how the potential landscape
can be manipulated by changing the 
relative strengths of the two recruitment processes, $\eta$.
This potential is directly defined as $V(m)=-\ln P_0(m)$,
where $P_0(m)$ is defined as the steady state probability distribution of $m$.
For simplicity the cell division is not taken into account here.
For small $\eta$, the valley at the M-state is much deeper than 
that of the A-state, and the landscape dramatically changes as $\eta$
increases to give a landscape where the A-state has a deeper valley than the M-state.

\section*{Discussion}
Positive feedback
in nucleosome modification 
is a powerful
mechanism to maintain
a dynamic bistable system, even with destabilizing
factors such as cell-division. Here we demonstrated how positive
feedback in itself can be reformulated into an epigenetic landscape with peaks and valleys that
reflects the underlying balance between feedback and noise. As long
as movements are small, dominated by single nucleosome
modifications, the movement in the landscape can be fully modeled
by a Langevin or Fokker-Planck equation with a first escape time calculated in analogy to Kramers.
When stochastic events are large, as during cell divisions, a transfer matrix method 
allowed us to extend the Fokker-Planck formalism and thereby to set a minimum 
timescale for the dynamics of a robust positive feedback.
Finally, we studied how the landscape could be 'tilted' by asymmetry
in the nucleosome modification reactions.

We expect that the transfer matrix method can be extended relatively easily to include other
nucleosome modification schemes, for example the 3-state model 
of \cite{dodd}. 
This approach is more powerful than the mean field approach
\cite{david-rus} in the sense that it allows one to explore 
the probability distribution.
The method can also deal with cases where the epigenetic system 
does not conform to a potential energy surface.
Noise associated with cell divisions can for example be included in systems with double negative feedback between
repressors, such as the CI-Cro feedback 
loop in the lysis-lysogeny switch of phage
lambda \cite{aurell}.

Our results are consistent with recent observations 
in mammalian cells in which increased cell division 
rates accelerated stochastic transitions between epigenetic 
states \cite{Hanna}. 

Epigenetic landscapes present a particularly appealing way to
discuss multi-stability of expression states in living systems.
The presented coupling between positive feedback and the
possibility for a drift in a landscape, may be useful for understanding
cases where bistable decisions are delayed,
as often seems to be the case in development.
Some epigenetic landscapes may define the activity of transcription factors
that act as histone modifying complexes,
and thereby subsequently define the input parameters for other
landscapes further down along a developmental pathway.
Thereby understanding of epigenetic stability and regulated tilting
of landscapes may speak to large classes of coupled switch systems.

\section*{Conclusion}
This paper explored theoretical implications of epigenetics as a 
dynamic phenomenon, where alternate 
states of gene expression are selected and maintained over multiple generations 
by on going dynamic processes.
Our approach builds on the assumption that the epigenetic 
states are maintained
through a positive feedback where nucleosomes of a certain kind recruit
enzymes which in turn convert other nucleosomes to the same kind.
By introducing a minimal model for such a dynamic system we 
demonstrated that the on-going nucleosome updating rates only 
 need 
to be $g\stackrel{>}{\sim} 10$ updates per nucleosome per generation 
to provide robust maintenance. This result is closely linked to the
convergence of the P(M) distribution after cell division (Fig. 4), 
which we find to be rather insensitive to the value of $F$.
Thus we expect that any model working with positive feedback
as a key maintenance factor, in principle would work with 
a moderate number of updates per generation. 

It is important to notice that the stability of the epigenetic 
state depends on the update rules at the cell-divistion especially 
in the two-state model. 
In our model, on average half of the nucleosomes come from parental DNA,
and the rest of them are either in the modified or the 
unmodified state with equal
probability, thus effectively 75\% nucleosome are in the same epigenetic
state as the parental DNA. 
Thus the most of the cells are still in the 
same valley in the epigenetic potential (Fig.2 g) and 
come back to the original epigenetic state. 
However, if half of the nucleosomes are replaced with unmodified 
nucleosomes at the cell division in the 2-state model, 
the system prefers unmodified states and one needs 
extra mechanisms to keep modified state stable \cite{david-rus}. 
It is not clear which mechanism operates at DNA replication
in different systems, but our formulation is applicable in 
both cases. More experimental information about the nucleosomes 
inserted after DNA replication will be critical in understanding 
the stability of epigenetic states.

Our analysis also showed formally
that bistability requires cooperativity of recruitment,
in the sense that eq. (\ref{eqPflux}) require more than first order terms in
order to provide separation between two states. 
In the model in Dodd \cite{dodd}
the recruitment was implemented by requiring that two independent
recruitment processes were of the same type, whereas we here
simply assumed that the two simultaneously recruiting
nucleosomes are of the same type.
Our analogy between system size and effective randomness allowed us to 
show formally that stability of inherited 
states will grow exponentially with the number
of nucleosomes in the considered region of the chromosome.
  
Our analysis opens for understanding development in epigenetic landscapes,
in terms of positive feedback mechanisms
that are linked to each other through expression of nucleosome
modifying enzymes.

\section*{Appendix}
\subsection*{Analytical calculation of the stability of a macroscopic
  state}
Assuming $V(m)$ is harmonic around $m=m_A$
and hence the initial distribution $P(m)$ is a Gaussian around 
$m=m_A$ (and $P(m_M)\approx 0$), we get
\begin{equation}
\left[P(m)\exp(V(m)) \right]_{m_A}^{m_M}\approx 
-P(m_A)\exp(V(m_A))=-\frac{1}{\sqrt{2\pi \sigma_A^2} }\exp(V(m_A)),
\end{equation}
with $1/\sigma_A^2=\left(d^2 V(m)/dm^2\right)_{m=m_A}$.
From the condition $dV(m)/dm|_{m=m_A}=0$ and eq. (\ref{eqV}), we have 
$m_A=1/2-\sqrt{1/4-1/F+1/(2N)}$,  and
\begin{equation}
\sigma_A^2=\left(\frac{d^2 V(m)}{dm^2}\right)_{m=m_A}^{-1}
=\frac{4N-F}{4N(FN+2F-4N)}
\approx\frac{1}{N(F-4)}
\approx\frac{1}{NF},
\end{equation}
where we assume $N>>1$ and $F>>1$ (i.e. $\alpha\approx 1$).

Approximating $V(m)$ as harmonic around the 
transition state T with $m=m_T =1/2$ 
and noting $V(m)$ takes the maximum at $m=m_T$, we have
\begin{eqnarray}
\int_{m_A}^{m_M} \frac{1}{D(m)} \exp(V(m)) dm
& \approx& \frac{1}{D(m_T)}\exp(V(m_T))\int_{-\infty}^{\infty} 
\exp\left[-\frac{(m-m_T)^2}{2\sigma_T^2}\right]dm\nonumber \\
&=&\frac{\sqrt{2\pi \sigma_T^2}}{D(m_T)}\exp(V(m_T))
\end{eqnarray}
with
\begin{equation}
\sigma_T^2=-\left(\frac{d^2 V(m)}{dm^2}\right)_{m=m_T}^{-1}
=-\frac{1+F/4}{(4-F)N-2F}
\approx\frac{1+F/4}{N(F-4)}
\approx\frac{1}{4N}
\end{equation}
Noting 
\begin{equation}
D(m_T)=\frac{\alpha(1/4+1/F)}{2N^2}
\approx \frac{1}{8N^2},
\end{equation}
we get
\begin{eqnarray}
\tau=\frac{1}{|J|}&\approx &\frac{8N^2}{\alpha}\cdot
\sqrt{2\pi}\sigma_T e^{V(m_T)}\cdot
\sqrt{2\pi}\sigma_A e^{-V(m_A)}\\
&\approx&
4\pi N\sqrt{\frac{4}{F}} \exp\left[V(m_T)-V(m_A)\right].
\end{eqnarray}
\section*{Acknowledgments}
This work was supported by the Danish National
Research Foundation.

\bibliography{histone}

\begin{thebibliography}{10}

\bibitem{grunstein}
M.~Grunstein.
\newblock Yeast heterochromatin: regulation of its assembly and inheritance by
  histones.
\newblock {\em Cell}, 93:325--328, 1998.

\bibitem{turner}
B.M. Turner.
\newblock Histone acetylation as an epigenetic determinant of long-term
  transcriptional competence.
\newblock {\em Cell. Mol. Life. Sci.}, 54:21--31, 1998.

\bibitem{felsenfeld}
G.~Felsenfeld and M.~Groudine.
\newblock Controlling the double helix.
\newblock {\em Nature}, 421:448--453, 2003.

\bibitem{dodd}
I.B. Dodd, M.A. Micheelsen, K.~Sneppen, and G.~Thon.
\newblock Theoretical analysis of epigenetic cell memory by nucleosome
  modification.
\newblock {\em Cell}, 129:813--822, 2007.

\bibitem{sedighi}
M.~Sedighi and A.~M. Sengupta.
\newblock Epigenetic chromatin silencing: bistability and front propagation.
\newblock {\em Phys. Biol.}, 4:246--255, 2007.

\bibitem{sneppen08}
K.~Sneppen, M.~A. Micheelsen, and I.~B. Dodd.
\newblock Ultrasensitive gene regulation by positive feedback loops in
  nucleosome modification.
\newblock {\em Mol. Sys. Biol.}, 4:182, 2008.

\bibitem{waddington}
C.H. Waddington.
\newblock {\em The strategy of the Genes; a Discussion of some aspects of
  theoretical biology}.
\newblock London: Allen \& Unwin, 1957.

\bibitem{goldberg}
A.D. Goldberg, C.D. Allis, and E.~Bernstein.
\newblock Epigenetics: A landscape takes shape.
\newblock {\em Cell}, 128:635--638, 2007.

\bibitem{thon}
G.~Thon and T.~Friis.
\newblock Epigenetic inheritance of transcriptional silencing and switching
  competence in fission yeast.
\newblock {\em Genetics}, 145:685--696, 1997.

\bibitem{rusche}
L.N. Rusche, A.L. Kirchmaier, and J.~Rine.
\newblock The stablishment, inheritance, and function of silenced chromatin in
  saccharomyces cerevisiae.
\newblock {\em Annu. Rev. Biochem.}, 72:481--516, 2003.

\bibitem{naar}
A.M. Naar, B.D. Lemon, and R.~Tjian.
\newblock Transcriptional coactivator complexes.
\newblock {\em Annu. Rev. Biochem.}, 70:475--501, 2001.

\bibitem{ruthenburg}
A.J. Ruthenburg, C.D. Allis, and J.~Wysocka.
\newblock Methylation of lysine 4 on histone h3: intricacy of writing and
  reading a single epigenetic mark.
\newblock {\em Mol. Cell}, 25:15--30, 2007.

\bibitem{kramer}
H.~A. Kramers.
\newblock Brownian motion in a field of force and the diffusion model of
  chemical reactions.
\newblock {\em Physica}, 7:284--304, 1940.

\bibitem{david-rus}
D.~David-Rus, S.~Mukhopadhyay, J.~L. Lebowitz, and A.~M. Sengupta.
\newblock Inheritance of epigenetic chromatin silencing.
\newblock {\em J. Theor. Biol.}, 258:112--120, 2009.

\bibitem{annunziato}
A.T. Annunziato.
\newblock Split decision: what happens to nucleosomes during dna replication?
\newblock {\em J. Biol. Chem.}, 280:12065--12068, 2005.

\bibitem{aurell}
E.~Aurell and K.~Sneppen.
\newblock Epigenetics as a first exit problem.
\newblock {\em Phys. Rev. Lett.}, 88:48101, 2002.

\bibitem{Hanna}
J.~Hanna, K.~Saha, B.~Pando, J.~van Zon, C.~J. Lengner, M.~P. Creygton, A.~van
  Oudenaarden, and R.~Jaenisch.
\newblock Direct cell reprogramming is a stochastic process amendable to
  acceleration.
\newblock {\em Nature}, 462:595, 2009.

\end{thebibliography}
\section*{Figure legends}

\begin{figure}[!h]
\includegraphics[width=0.65\textwidth]{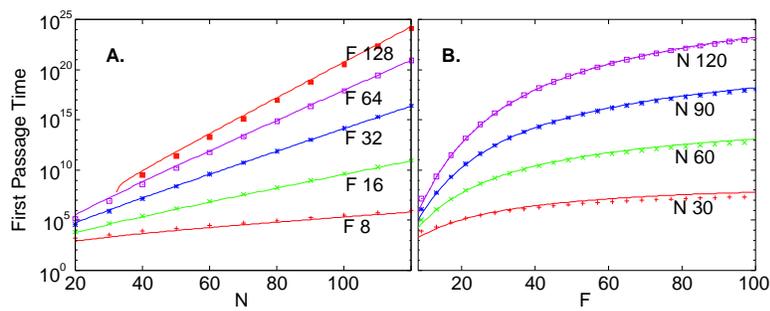}
\caption{{\bf Stability of a macroscopic state.}
The first passage time, or the average life time 
of an epigenetic state $\tau$, is shown as a function of $N$ (A)
and $F$ (B). 
Symbols show numerical results and lines show
analytical results from eq.~(\ref{Kramer}). 
Time is counted in units of attempted nucleosome
updates per nucleosome. \label{fig4} }
\end{figure}

\begin{figure}[!h]
\includegraphics[width=0.65\textwidth]{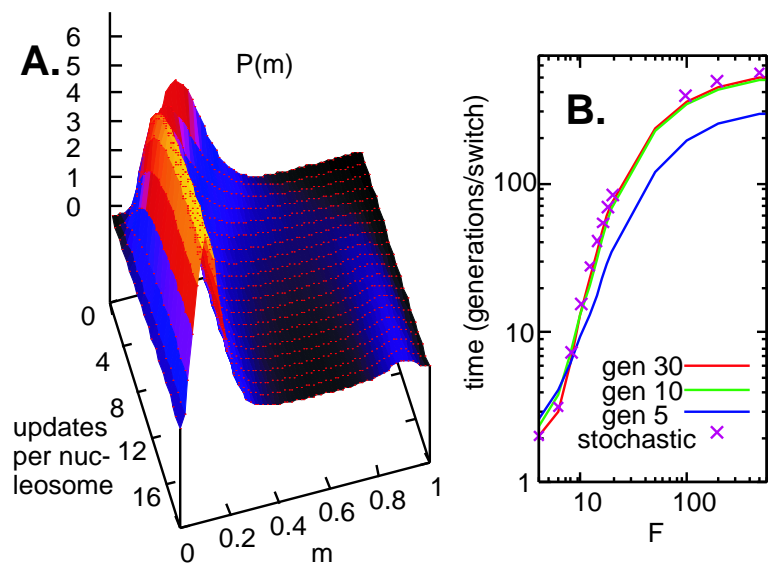}
\caption{{\bf The switching of the 
macroscopic state via randomization at the cell division.}
(A) The averaged development of
the $M$ distribution over one cell generation, with
``time'' measured in attempted updates per nucleosome. 
The simulation
uses $F=10$ and a generation length of 20 updates
per nucleosome. The evolution starts just after a cell division,
where a randomization (using eq.~\ref{celldiv}) 
is followed by a drift imposed
by the epigenetic landscape. 
After about 10 updates one sees that the $P(M)$ reaches a nearly
stationary distribution, where a fixed fraction has switched to the
alternate state. 
Just before the next cell division one
resets $P(M)=0$ for $M>N/2$, and
renormalizes the distribution. 
Iterating this process, panel B
shows the average number of generations 
needed before escape
(see escape time discussion in the text).
A direct Monte-Carlo simulation result
with a generation time of 30 is also shown by symbols.
}
\label{matrix}
\end{figure}

\begin{figure}[!h]
\includegraphics[width=0.65\textwidth]{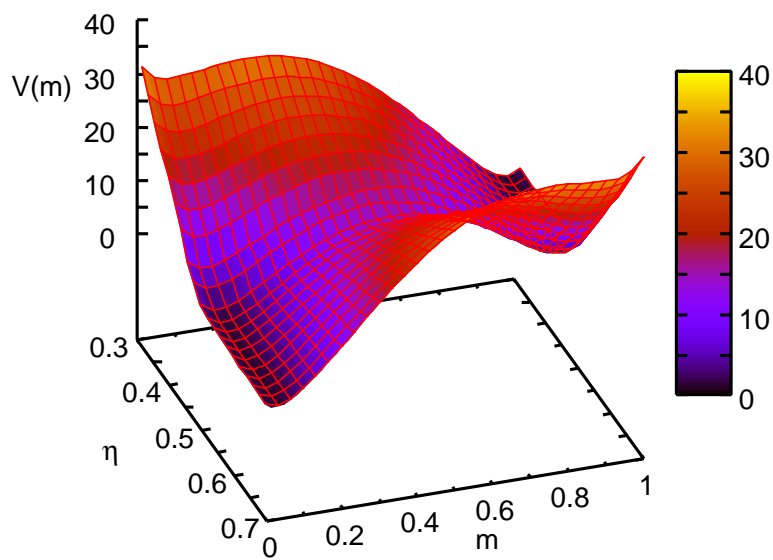}
\caption{{\bf Epigenetic landscape with asymmetry.}
The figure shows $V(m)=-\ln(P_0(m))$ obtained from
the model with 
asymmetric transition rates
eqs.~(\ref{eqPfluxtilt}) using the transfer matrix method.
For simplicity the cell division is not taken into account here.
The figure shows how the landscape gradually changes from monostable, to
 bistable
and then again to monostable as one changes the parameter eta.
This ``asymmetry'' parameter $\eta$ would in principle be under 
biological control through changes of concentrations of modifying 
enzymes associated with the recruitment processes.
The total recruitment to noise ratio is fixed at $F=20$. 
The figure illustrate a dramatic
tilting of the epigenetic landscape by a moderate change in
nucleosome modification rates. }\label{matrixx}
\end{figure}
\end{document}